\documentclass[12pt]{article}
\usepackage[colorlinks]{hyperref}
\usepackage{graphicx}
\usepackage{color}
\begin{document}
\thispagestyle{empty}
\renewcommand*{\appendixname}{Appendix}
\newcommand{\fr}{``flux rule''}
\begin{center}
{\Large\bf Teaching Electromagnetism in elementary physics or upper high school courses}\\
\vskip4mm {Biagio Buonaura  and Giuseppe Giuliani}\\
\vskip4mm{\small Formerly, ISIS Albertini, Via Circumvallazione
292, 80035 Nola, Italy\\ Formerly,  Dipartimento di Fisica, Universit\`a
degli Studi di Pavia, via Bassi 6, 27100 Pavia, Italy}\\
{\verb"bbuonaura@gmail.com"}\hskip1cm {\verb"giuseppe.giuliani@unipv.it"}

\end{center}
{\bf Abstract.}
				{Traditionally, Electromagnetism is taught following the
chronological development of the matter. The final product of this
path is a presentation of Electromagnetism realized by adding one
layer over another with the risk of transferring concepts and
formulae from Electrostatics to Electrodynamics. In this paper, we
suggest a new approach based on the idea that the matter should
be presented within the conceptual framework of
Maxwell-Lorentz-Einstein Electromagnetism. This approach is
founded on the concept of a field as a primary theoretical entity
and on the statement that a point charge produces, in general, an
electric and a magnetic field and that the force exerted by these
fields on a point charge is the Lorentz's force. Developing this
idea, one finds that macroscopic laws corroborated by
experiments have a microscopic origin. It also follows that the
electromotive force induced in a closed conducting circuit must
be defined as the line integral of Lorentz's force on a unit positive
charge. This definition leads to a local law of electromagnetic
induction, Lorentz's invariant for rigid and filiform circuits. This
law contrasts with what Feynman labeled as the \fr -- generally
taught in textbooks and teaching practices -- downgrading it from
the status of physical law. Particular attention is given to the
teaching dilemma of Maxwell's equations: ignore them, write them
in integral form, or speak of them, focusing on their conceptual
and physical meaning.}
\section{Introduction}
Traditionally, in elementary physics and upper high school courses, Electromagnetism is
taught following the chronological development of the matter \footnote{This approach is typical for first-level university
textbooks.}. {With the term ``elementary physics courses,'' we refer to college courses whose level is intermediate between high school and university. As for high schools, the students' mathematical and physical background knowledge varies widely from country to country. As for Italy, we refer to the last three years of scientific Lyceum.}
\par
The first argument is Electrostatics, based on
Coulomb's law. \label{conc_campo_0}Usually, the definition of the electric field derives
from the concept of force. There are two typical approaches. One
states the Coulomb law as:
\begin{equation}\label{coul_2}
F=q\left(\frac{kQ}{r^2}\right),
\end{equation}
where $k$ is a  constant depending on the unit system employed,
$Q$ is a large charge, and $q$ is a small enough charge called the
test charge. The term in brackets depends only on the charge $Q$.
Then, the norm of the electric field vector is defined as $F/q$, and
$\vec E=\vec F/q$ gives the electric field vector. This approach is
found, for instance, in \cite[p. 46]{ppc}.
\par
Otherwise, one can define the electric field ``as the force exerted
on a unit positive charge by a charged body''. See, for instance,
\cite[p. 421]{pssc}. If we leap about fifty years, we see that nothing
has changed \cite[pp. 630 - 635]{halres}, \cite[pp. 450
451]{giancoli}. In these definitions the field concept is derived
from that of force: it is not a primary concept.
\par
{The further development of the matter deals with continuous
currents. The microscopic nature of currents is somewhat
remembered; however, it generally does not enter into any
calculation. Here, we have a conceptual discontinuity. While
Coulomb's law speaks of point charges, their constitutive role in
currents is overlooked.} \par The magnetic field is introduced by
considering magnets or by recalling {\O}rsted's discovery of the
deviation of a magnetized needle by a current-carrying wire. {The
experiments by Amp\`ere allow us to discuss the magnetic forces
exerted by current-carrying wires within a macroscopic
description. The role of moving point charges as ultimate sources
of the magnetic field needs to be recovered.}
\par
The magnetic component of Lorentz's force is introduced as an
experimental finding. Then, the force exerted by an
electromagnetic field on a point charge becomes:
\begin{equation}\label{lorfor}
\vec F=q(\vec E+\vec v\times\vec B),
\end{equation}
where $\vec v$ is the charge's velocity. Nonetheless, the induced
electromotive force is defined as: \begin{equation} \mathcal E=
\oint \vec E\cdot\vec {dl}, \label{emfstatic}\end{equation} instead
of:
\begin{equation} \mathcal E= \oint (\vec E+ \vec v_c \times \vec
B)\cdot\vec {dl},\label{emfdynamic}\end{equation} as the expression of
Lorentz's force implies and as argued in \cite{epl,ajp} ($\vec v_c$
is the charge's velocity).
\par
{Critical issues are constituted by
electromagnetic induction and Maxwell's equations. For
electromagnetic induction, textbooks, and teaching practices rely
on the \fr. Feynman \cite[pp. 17.1-17.3]{feyn2} and, more recently,
Giuliani \cite{ajp} have shown that the \fr is only a calculation
shortcut and not a physical law.} As for Maxwell's equations, their
differential form requires mathematical skills that are out of reach. {Then, how to manage these two fundamental
issues?}
\par
An epistemological stand accompanies this traditional approach,
according to which experiments must induce physical laws.
Moreover, the diffuse habit of referring to the student's daily life
and sensorial experiences obscures the role of the theories and
the need for abstraction. The final product of this traditional path
is a presentation of Electromagnetism realized by adding one layer
over another with the risk of transferring concepts and formulae
from Electrostatics to Electrodynamics. The definition of the
induced electromotive force given by Eq. (\ref{emfstatic}) instead
of the correct one (\ref{emfdynamic}) is a striking  example.
\par
{The broad literature in Physics Education generally deals with the
students' difficulty understanding fundamental concepts or
focuses on typical students' misconceptions or
misunderstandings. These studies often rely on multiple-choice
tests, sometimes integrated with interviews. The validity of
multiple-choice tests as a mean to evaluate the student's
understanding have been studied by many authors. See, for
instance, \cite{stem} and the references therein. However, their
utility in unrevealing students' misconceptions or
misunderstandings seems out of doubt.}
\par
{We shall discuss two papers particularly suited for our discourse.
Sa\u{g}lam and Millar used a multiple choice test administered to
English and Turkish upper high school students \cite{saglam}.
They also interviewed a sample of Turkish students to determine
the reasoning followed in answering the test. The questions
concerned three fundamental topics of Electromagnetism:
``magnetic field (caused by moving charges), magnetic forces (on
moving charges and current-carrying wires), and electromagnetic
induction.'' \cite[p. 546]{saglam}. Sa\u{g}lam and Millar found four
types of difficulty in understanding these electromagnetic topics
\cite[p. 558]{saglam}:
\begin{enumerate}
\item inappropriate analogies between the effect of magnetic
and electric field on electric charges
\item an over-literal flow interpretation of magnetic field lines
\item incorrect use of direct cause-effect reasoning in situations
where it does not apply
\item confusion between change, and rate of change, of
variables (such as magnetic flux).
\end{enumerate}
In their conclusions, Sa\u{g}lam and Millar write \cite[p. 564; our
italics]{saglam}:
\begin{quote}\small
By using samples from two countries, the study also shows a
striking level of agreement in the questions (and hence perhaps
the ideas) that students found most straightforward and most
difficult. {\em This increases confidence that learning difficulties
are due to inherent characteristics of the material, rather than
stemming from the way it is taught (which is quite different in the
two countries)}.
\end{quote}
Indeed, Maxwell's Electromagnetism, in its modern version --
owing to the contributions by Lorentz and Einstein
(Maxwell-Lorentz-Einstein Electromagnetism - MLE), is
conceptually challenging for two fundamental reasons: it requires
a theoretical approach centered on the concept of the field; it
incorporates the special relativity result of a limited speed for
material particles and physical interactions. These conceptual
features are at stake with the Newtonian view, where
forces-at-a-distance are the main actors, and every speed is
allowed. Coulomb's law operates in a strictly Newtonian view. We
suggest that the student's difficulties are enhanced by teaching
practices -- fueled by textbooks and syllabuses - based on the
chronological development of the matter. A possible way out is
teaching Electromagnetism within the MLE conceptual
framework.}
\par
{A relatively recent study by Zuza et al. \cite{zuza} reinforces this
working hypothesis. The authors used a test constituted by six
conceptual free-response questions proposed to first or
second-year University students in three different European
countries (Spain, Belgium, and Ireland). In this study also, the
student's difficulties are independent of their country, ``regardless
of differences in their educational system and cultural
background''. The fact that people involved in the test were first
years University students is not significant unless we assume that
the misconceptions or misinterpretations surfaced were due to the
University's teaching and that this teaching has completely
canceled previous misconceptions or misinterpretations. After a
careful discussion of the answers, the authors write:
\begin{quote}\small
In conclusion, we believe that more attention should be paid to the
specific characteristics of field theory and the difference between
fields and forces, with particular emphasis on the conceptual
interpretation of the interaction process rather than rules. Such an
approach would guide students in the transition from a Newtonian
to a Maxwellian viewpoint, underpinned by a changing view of the
field from a calculational convenience to a physical entity.
\end{quote}
The difficulty of substituting the Newtonian force-centered
viewpoint with the field conceptual framework demands a change
also in how we teach electromagnetic phenomena in high school or in elementary physics courses.}
\par
{The proposal discussed in this paper requires abandoning the
chronological development of the matter and  presenting
Electromagnetic phenomena to students within the
conceptual framework of MLE. }
\par
{This paper is organized as follows. Section \ref{proposal}
presents the main traits of our proposal. Section \ref{field} deals
with the concept of the field as a primary theoretical entity. Section
\ref{moving} suggests how to introduce the idea that a point
charge produces, in general, an electric and a magnetic field.
Section \ref{vectorp} deals with the opportunity of introducing the
vector potential in an elementary way. Section \ref{emi} treats the
problem of electromagnetic induction. Section \ref{what}
discusses to what extent, if any, instructors should speak about
Maxwell's equations. The reader will find a general discussion and
conclusive remarks in the last section \ref{discon}.}
\section{The broad lines of the proposal}\label{proposal}
{The present proposal is not -- and could not be -- a receipt ready to use.
It only highlights the main and interrelated conceptual features of MLE that could be transferred into high school or elementary physics teaching. It can be read at two levels: as an occasion for refreshing the instructors' cultural background or as a guide for trying some changes in the teaching practices.
We know the many constraints that limit the teachers's creativity (at least in Italy): the syllabuses prepared by the Ministry of
Education and the local tendency to standardize teaching
practices in all the classrooms, with the adoption of the same
textbooks, also in the view of preparing the students for the
final state exam.
Also, the textbook publishers' policy of supplying many developed didactic tools (lessons included) does not stimulate instructors's creativity.
The proposal contemplates the possibility of using some formalism more complicated than the one commonly used. However, at each step of this kind, it is stressed that the important thing is the concept, not the formula accompanying it. The choice of formalism is left to the teacher, who must consider his teaching context} \footnote{{The physical and mathematical background knowledge of the students
potentially involved in the experimentation varies widely from
country to country. The instructors must decide what
the extent to adapt the present proposal to their teaching
conditions.}}.
\par
An introductory discourse should take up again the difference
between Galileo's and Einstein's relativity principles. Both
principles require that physical laws have the same form in every
inertial frame. However, while the former obeys Galileo's
coordinates transformations, the latter obeys Lorentz's. The
former allows physical interactions with infinite speed. Instead, the
latter implies that physical interactions can propagate only with a
finite velocity whose upper limit is light's speed in a vacuum. The
physical differences between the two views are well illustrated by
considering the gravitational field produced by a mass in
Newtonian mechanics and the electric field produced by a point
charge in Electromagnetism. In Newton's gravitational theory, the
gravitational field produced at the point $\vec r_1$ at the time $t$
by a mass depends on the position $\vec r_2$ of the mass at the
same time $t$ [physical interactions propagate at infinite speed].
{Instead, in Electromagnetism, in a chosen inertial reference frame,
the electric field produced at the point $\vec r_1$ at the time $t$ by
a moving point charge depends on the position of the charge at an
earlier time, named `retarded time': to this retarded time also
pertain a `retarded position', a `retarded velocity', and a `retarded
acceleration' of the point charge.}
\par
{Independently of the fact that instructors use Maxwell's equations in some form or not, they should emphasize that these equations
have allowed,
through the work of Hertz, Lorentz, and Einstein, the possibility of
describing all electromagnetic phenomena in a vacuum in an
axiomatic way. This step is, among others, necessary for correcting the students' conception of physics, and in general of science, as a discipline founded essentially (if not only) on experiment. Consequently, teachers should talk about
physicists' two principal methods to establish their discipline's
laws: the inductive and the axiomatic method. This discourse should conveniently refer to the
historical development of Electromagnetism.
The inductive method was dominant during the nineteenth
century. Referring to Faraday's fundamental contributions to
electromagnetic induction, Maxwell wrote:  ``The method which
Faraday employed in his researches consisted in a constant
appeal to experiment as a mean of testing the truth of his ideas,
and a constant cultivation of ideas under the direct influence of
experiment'' \cite[p. 163]{treatise2}. Maxwell wrote his equations
after having derived many laws from experimental results. Hertz,
referring to Maxwell's equations, vindicated the importance of the
axiomatic method with these words:}
\begin{quote}
These statements [Maxwell's equations] form, as far as the ether is
concerned, the essential parts of Maxwell's theory. Maxwell arrived
at them by starting with the idea of action-at-a-distance and
attributing to the ether the properties of a highly polarisable
dielectric medium. We can also arrive at them in other ways. {\em
But in no way can a direct proof of these equations be deduced
from experience. It appears most logical, therefore, to regard
them independently of the way in which they have been arrived at,
to consider them as hypothetical assumptions, and to let their
probability depend upon the very large number of natural laws
which they embrace \cite[p. 138, italics added]{ew}.}
\end{quote}

Instructors should also draw students' attention to their study of Newtonian mechanics and thermodynamics within an axiomatic approach.  
\par
The present proposal rests on three cornerstones:
{\begin{enumerate}
\item The use of the field concept as a primary theoretical entity
and the necessity of introducing the field concept beginning
with the gravitational interaction (In Italian Scientific Lyceums, during the third year).\label{conc_campo}
\item The idea that electromagnetic phenomena must be treated
within the conceptual domain of Maxwell-Lorentz-Einstein
Electromagnetism.\label{mle}
\item The statement that only local equations can be
{\em interpreted}
causally. This statement stems from special relativity, and it
means that an equation is local if it connects two physical
quantities at a given point at the same time $t$, or the equation
connects a physical quantity at point $\vec r_1$ at the time
$t_1$ to another physical quantity at the point $\vec r_2$ at the
time $t_2$, with $t_2>t_1$, provided that the distance between
the two points $\le c(t_2-t_1)$.\label{local}
\end{enumerate}
}
{Point \ref{conc_campo} above is the more delicate because it involves the passage from the action-at-a-distance view to that of the field. The study by Zuza et al. \cite{zuza}, discussed in the Introduction, shows how this passage disorients university students of the first two years. We suggest that the traditional way to introduce the field presented by textbooks (see page \pageref{conc_campo_0}) does not help this conceptual transition and that a  new approach is necessary. In the next section, we shall see how a  passage from Feynman's {\em Lectures} can help us.}
{Point  \ref{mle} leads to the introduction, from the beginning, of the idea that a point charge produces, in general, an electric and a magnetic field and that, coherently, the force exerted by these fields on a point charge is the Lorentz's force. Point \ref{local} directly impacts the treatment of electromagnetic induction, the relation between the electric and the magnetic field during their propagation, and their causal connection with the sources (moving electric charges).
Finally, we emphasize that the formulae are written as concisely as possible in the following sections. The instructors should adapt them to their teaching context.
}
\subsection{The field concept}\label{field}
Introducing the concept of the field as a primary theoretical entity
needs some practice of abstraction. As Feynman put it \cite[p.
15-7]{feyn2}\label{ft}:
\begin{quote}\small
What we mean here by a field is this: a field is a mathematical
function we use for avoiding the idea of action at a distance. If we
have a charged particle at the position $P$, it is affected by other
charges located at some distance from $P$. One way to describe
the interaction is to say that the other charges make some
``condition'' -- whatever it may be -- in the environment at $P$. If
we know that condition, which we describe by giving the electric
and magnetic fields, then we can determine completely the
behavior of the particle -- with no further reference to how those
conditions came about.
\par
[\dots]
\par
A field is then a set of numbers we specify in such a way that what
happens {\em at a point} depends only on the numbers {\em at that
point}. We do not need to know any more about what's going on at
other places.\footnote{The original text speaks of a ``real'' field. We
have omitted the adjective ``real'' because its use by Feynman
concerns the epistemological status of the vector potential.
Indeed, Feynman acknowledges from the beginning that ``First we
should say that the phrase ``a real field'' is not very meaningful''.}
\end{quote}
Feynman's conception of the electromagnetic field stresses that it
is only a theoretical tool for describing electromagnetic
phenomena, with no commitment to its existence in the world.
Indeed, a theory aims to predict the values that the physical
quantities can assume. Its aim is not to describe what is
happening in the world: we do not have any means to ascertain
that. This conception of the theories and their fundamental role
contrasts textbooks' inductive and naively realistic stand.
\par
{According to point  \ref{conc_campo} above, the field concept must be
introduced in the physics course as soon as possible. The
occasion is, naturally, Newton's gravitational law. Besides the
traditional equation in terms of the force of attraction between two
(point) masses, instructors should introduce the description in
terms of the gravitational field. A mass $M$ produces at the point
$\vec r$ a gravitational field $\vec g$ given by:
\begin{equation}\label{gravcampo}
\vec g= -G \frac{M}{r^3}\vec r,
\end{equation}
where $G$ is the gravitational constant. The field $\vec g$ is such
that, if another mass $m$ is positioned at the point $\vec r$, then a
force $F$ given by:
\begin{equation}\label{gravforza}
\vec F = m \vec g,
\end{equation}
acts on the mass $m$. Two points should be emphasized: A) the
gravitational field has the dimensions of an acceleration and B) the
conceptual scheme is: mass $\rightarrow$ field $\rightarrow$ force
on another mass. Instructors should comment on how deeply the
description in terms of field differs from that of the
action-at-a-distance. Moreover, point A) suggests a series of
reflections that could be developed according to the teaching
context. See, in the Appendix, the section
\ref{gravfield}.}
\par
{Similarly, the treatment of Galilean-Newtonian relativity is the occasion of introducing the basic concepts of Einstein's relativity and discussing their fundamental differences. Instructors could do this based on the following points:}
{\begin{enumerate}
\item In an inertial reference frame, the acceleration measured by an accelerometer is null (see, for details, the section \ref{gravfield} in the Appendix). Accordingly, this is the best way of defining an inertial reference frame.
\item In both Galilean-Newtonian relativity and Einstein's, all phenomena develop in the same way in every inertial frame, i.e., the equations describing each phenomenon have the same form in every inertial frame.
\item In both cases, the space is Euclidean, i.e., it is homogeneous and isotropic, and the (variable) time is homogeneous.
\item The difference between the two approaches lies in that, while in the Galilean-Newtonian case, the coordinates' transformations are the so-called Galilean transformations, in Einstein's, the so-called Lorentz transformations are valid. The latter introduces the big novelty of a speed limit, given by the speed of light in a vacuum. This speed limit is responsible for the time-dilation and the length-contraction effects, as it can be easily seen by putting in their formulae $c=\infty$, passing, in this way, from Lorentz's to Galilean-Newtonian transformations. In \cite[chapter II]{erq}, \cite[chapter III]{ugarov}, and \cite{bondi}, instructors will find derivations of the basic formulae of Einstein's kinematics obtained with thought experiments with the exchange of light pulses of ideal null duration between two inertial reference frames. The mathematics involved are elementary algebraic calculations.
\end{enumerate}
}
\subsection{The electromagnetic field produced by a moving charge}\label{moving}
Treating electromagnetic phenomena within the conceptual
framework of MLE requires a microscopic description of the
phenomena. This description, in turn, implies that instructors
must -- as an introductory but fundamental part -- give a picture of
what matter is made of. How detailed this picture can be, depends
primarily on the teaching context. This description should include
information on what atoms and molecules are made of and how
atoms enter and behave in conducting or insulating material. A
focus should be put on the conducting mechanism in metals and
the fact that electric currents in metals are made of moving
electrons. Without ignoring that, by convention, mobile electric
carriers are considered positive. The equation of the current
density vector:
\begin{equation}\label{densitacorrente}
\vec J =nq\vec v_q,
\end{equation}
where $n$ is the number of charges $q$ per unit volume and $\vec
v_q$ their velocity, should be written explicitly. From Eq.
(\ref{densitacorrente}), it follows that if the charge $q$ is that of
the electron, then the direction of the vector current density is
opposed to that of the electrons' velocity. The electric current
through a surface $S$ is then defined as:
\begin{equation}\label{current}
I=-\int_S \vec J_e\cdot\hat n\, dS,
\end{equation}
where $\vec J_e$ is the electrons' current density, $\hat n$ is the
unit vector perpendicular to the surface element $dS$, and where
we have taken into account the convention about the current's
charge carriers. Instead of Eq. (\ref{current}), instructors can use
the simplified version in which the surface $S$ is perpendicular to
the motion of the charges. This simplified treatment is particularly
apt in the case of a metal wire. Considering this case, the stress
must be on the charge velocity being the drift velocity.
\par
Within this conceptual framework, instructors can state that, in
general, an electric charge produces an electric field $\vec E$ and
a magnetic field $\vec B$ and that an electromagnetic field exerts
on a point charge $q$ a force that is given by:
\begin{equation}\label{lorentz}
\vec F= q(\vec E+\vec v_q\times\vec B),
\end{equation}
where $\vec v_q$ is the velocity of the charge. Eq. (\ref{lorentz}) is
named ``Lorentz force''. \footnote{Indeed, as shown in \cite{ajp,
max_598}, Maxwell anticipated the expression of the Lorentz
force.}
\par
These statements can be grounded on experimental observations.
In a vacuum, the electric field produced at the point $\vec r$ by a
charge {\em at rest} ($\vec v_q=0$) at the origin has been proved
to be:
\begin{equation}\label{coulombfield}
\vec E=\frac{1}{4\pi\varepsilon_0}q \frac{\vec r}{r^3},
\end{equation}
where $\varepsilon_0$ is the dielectric constant in a vacuum.
Equation (\ref{coulombfield}) has been corroborated by
experiments with the Cavendish method \cite{caven}. Precisely,
this method tests the formula:
\begin{equation}\label{caven}
E=\frac{1}{4\pi\varepsilon_0}\frac{q}{r^{(2\pm \alpha)}},
\end{equation}
with the aim of reducig the value of $\alpha$ as much as possible.
Cavendish obtained $\alpha \le 2\times 10^{-2}$; Maxwell improved
to $\alpha\le 1/21600\approx 4.6\times 10^{-5}$ \cite[p.
77]{treatise1}. Modern measurements have reduced the value of
$\alpha$ to about $10^{-17}$ \cite{cinesi}.
\par
In discussing Cavendish's method, instructors should stress the
relevance of continuously increasing the accuracy of our
knowledge of fundamental physical laws and constants. They
should also underline that it is based on an axiomatic approach.
Indeed, the inverse square law is assumed to be true, and its
implication -- the inside conducting sphere must be free of
charges -- is tested by experiment. This axiomatic approach
should be compared with Coulomb's inductive experiment to
underlying the variety of methods used by physicists to unveil the
properties of phenomena.
\par
As for the magnetic field, the issue is more delicate. We could
begin by recalling {\O}rsted's experiment on the magnetic effect of
a continuous current. Since macroscopic currents in metals are
made of electrons moving with constant velocity (drift velocity),
we can assume that a moving charge produces a magnetic field.
Since the expression of the magnetic field can be obtained only by
fully developing the implications of the modern formulation of
Maxwell's equation in a vacuum, we can only state that the
magnetic field produced by a moving charge is given by:
\footnote{For the calculation of the electromagnetic field produced
by an arbitrarily moving charge, see, for instance, \cite[pp.
870-877]{andy}.}
\begin{equation}\label{Bfield}
\vec B \approx {{\mu_0}\over{4\pi}} q{{\vec v_q \times \vec r_{21}}
\over{r^3_{21}}},
\end{equation}
where $\vec v_q$ is the velocity of the charge $q$ and $\vec
r_{21}=\vec r_1-\vec r_2$ is the vector pointing from the position
$\vec r_2$ of the charge to the position $\vec r_1$ of the point in
which the field is calculated. The sign $\approx$ reminds us that
Eq. (\ref{Bfield}) is approximately valid if the velocity of the charge
$v_q\ll c$ and its variations are sufficiently slow to ignore the
acceleration effects. Within this approximation, the retarded
quantities of the charge $q$ (position and velocity) can be
replaced by the actual ones. The validity of Eq. (\ref{Bfield}) rests
on its experimental corroboration. Indeed, Eq. (\ref{Bfield}) can be
used to calculate the magnetic field produced by a continuous
current flowing in a long enough rectilinear wire or the magnetic
field produced by a continuous current flowing in a wire of
arbitrary form (Biot-Savart's law). These macroscopic equations
have been experimentally tested. Instructors could also add that --
in the same approximations of Eq. (\ref{Bfield}) -- the electric field
produced by a moving charge is:
\begin{equation}\label{Efield}
\vec E \approx {{q}\over{4\pi\varepsilon_0r^3_{21}}}\left(\vec r_{21} -r_{21}\frac{\vec v_q}{c} \right).
\end{equation}
By using the basic relation:
\begin{equation}\label{magneticoapp}
\vec B = \frac{1}{c }\left(\frac{1}{r_{21}^*}\vec r_{21}^*\times \vec E\right)\approx \frac{1}{c}\left(\frac{1}{ r_{21}}\vec r_{21}\times \vec E\right),
\end{equation}
where $\vec r_{21}^*$ is the retarded distance between the charge
and the point in which the field is calculated, one can obtain the
expression of the magnetic field (\ref{Bfield}). Going on, we should
develop some order of magnitude calculations. Let us consider a
long enough rectilinear metallic wire with a steady current. If the
wire has a section of a square millimeter and a current of one A
flows in it, the electron's drift velocity comes out to be $\approx
7.34\times 10^{-5}$ $\rm m s^{-1}$. Then, the second term of
equation (\ref{Efield}) is approximately $2. 44\times 10^{-13}$
smaller than Coulomb's term, and can be ignored in the
calculation of the electric field produced by the electron. However,
its presence is fundamental in calculating the magnetic field
produced by a slowly moving electron (Eq. \ref{Efield}).
\par
The instructors should adapt the above treatment to their teaching
context by keeping the essential concept: a moving charge
produces an electric and a magnetic field responsible for the
magnetic effects of the current flowing in a wire. {Here, there is an
intriguing problem. We have stated that the magnetic field is
produced by moving charges. Then, we learned that a moving
charge adds a correction to the value of the electric field produced
by the same charge at rest. Is there a physical quantity that can
describe both phenomena? Instructors know that this quantity
exists and is the so-called vector potential $\vec A$.}
\subsection{The vector potential}\label{vectorp}
{Instructors should say at least some words about the vector
potential to illustrate the conceptual role played by it. Students are
introduced from the beginning to the scalar potential $\varphi$.
Then, the idea that another potential exists should not appear as a
strange thing. The sources of the scalar potential are static
distributions of charges; the sources of the vector potential are
the currents, namely, charges in motion. From the knowledge of
its sources, i.e., charges in motion, we can calculate the value of
the vector potential $\vec A$. Knowing the vector potential, we can
calculate the magnetic field produced by the moving charges
through a relation involving a particular spatial variation of the
vector potential. Instructors know that this relation is ${\rm curl}
\vec A= \vec B$ (it is not necessary to show this equation to the
students). Moreover, a particular temporal variation of the vector
potential yields the contribution of the moving charges to the
electric field ($-\partial \vec A/\partial t= \vec E$). Then the
complete expression of the electric field is given by the sum of the
contributions from charges at rest and from charges in motion:
$\vec E=- {\rm grad}\varphi-\partial \vec A/\partial t$. Again, it is not
necessary to show this equation to the students. {In \cite[sec. VII]{ajp} the reader will find a detailed proposal for introducing the vector potential in elementary physics and high school courses.}} A less recent
proposal to introduce the vector potential in high
schools, can be found in \cite{giliberti}.
\subsection{Electromagnetic induction}\label{emi}
Textbooks and teaching practices describe electromagnetic
induction with what Feynman labeled as the \fr, downgrading it
from the status of physical law \cite[pp. 17.1-17.3]{feyn2}. The \fr
states that:
\begin{equation}\label{fluxrule}
\mathcal E= -\frac{d}{dt}\int_S \vec B\cdot \hat n \, dS= -\frac{d\Phi}{dt},
\end{equation}
where $\vec B$ is the magnetic field, and $S$ is any surface that
has the circuit as a contour. As shown in \cite{ajp}, the \fr is not a
physical law but only a calculation shortcut that must be handled
carefully. {Instead, the law of electromagnetic induction is founded
on the definition of the induced {\em emf} as \cite{epl,ajp}:
\begin{equation}\label{naturaldef}
\mathcal E=\oint_l (\vec E+\vec v_c\times\vec B)\cdot\vec{dl}= \oint_l \vec E\cdot\vec{dl}+ \oint_l  (\vec v_c\times\vec B)\cdot\vec{dl} ,
\end{equation}
where the electric field $\vec E$ and the magnetic field $\vec B$
are solutions of Maxwell's equation, and $\vec v_c$ is the velocity
of the positive charges that, by convention, are the current
carriers. This integral yields -- numerically --  the work done by the
electromagnetic field on a unit positive charge through the entire
loop. Eq. (\ref{naturaldef}) is local because it connects the physical
quantity $\mathcal E$ defined on the line $l$  at the time $t$ to
other physical quantities defined at every point of the line $l$ at
the same instant $t$.}
\par
{The expression of the electric field in Eq. (\ref{naturaldef})
contains a particular time dependence of the vector potential $\vec
A$ (its partial derivative with respect to time $-\partial\vec A/\partial
t$), as explained in section \ref{vectorp}. Then, the induced {\em
emf} is the sum of two line integrals as shown by the last equality
of Eq. (\ref{naturaldef}). The induced {\em emf} thus obtained
describes all known phenomena of electromagnetic induction
\cite{ajp}. See also, in the Appendix, the  section \ref{emisec}.}
\par
{Let us apply Eq.  (\ref{naturaldef}) to the relative inertial motion of
a magnet and a rigid, filiform circuit. For Einstein, this (thought)
experiment was one of the reasons for founding special relativity
\cite[Engl. transl., p. 140]{ein05r}. In the reference frame of the
magnet, there is no electric field. Therefore, only the second
integral of Eq. (\ref{naturaldef}) is operative. Instead, in the
reference frame of the circuit, both integrals are, in principle,
operative. However, the last integral of Eq. (\ref{naturaldef}) is null
because $\vec v_c=\vec v_d$ and $\vec v_d$ is always parallel to
$\vec {dl}$. The circuit sees the vector potential produced by the
magnet varying with time owing to the relative motion between the
magnet and the circuit. In conclusion: in the reference frame of the
magnet, only the magnetic component of Lorentz's force on a unit
positive charge is operative; in the reference frame of the circuit,
only the time variation of the vector potential operates.}
\par {It will be of great pedagogical value to experiment on this fundamental topic. An experiment of this kind has been described in detail in \cite{ggstud}. The laboratory session is held before any electromagnetic induction discourse but after the special relativity lessons. Students, divided into couples,  are invited to experiment at will. After about an hour or so of experimenting, students are asked to describe what they have seen with a formula. The instructor intervenes as little as possible. Spontaneously, the students describe the observed phenomena in the magnet reference frame. Then, the students are asked to describe their observations in the reference frame of the moving coil. After some discussion, the instructor suggests to guess a formula that obeys the locality principle. In this way, students learn or apply the principle that the equation describing a phenomenon must have the same form in every inertial frame.  }
\par
{It is possible to rewrite Eq. (\ref{naturaldef}) in terms of a single
surface integral under severely restricting conditions concerning
the integral \footnote{{The following calculations are for the instructors. Considering the available mathematical tools, they should adapt them to their teaching context.}}:
\begin{equation}\label{resticted}
\oint_l  (\vec v_c\times\vec B)\cdot\vec{dl}.
\end{equation}
Let us consider a rigid and filiform circuit that moves with velocity
$V$ in the laboratory. Let us further assume that the motion of the
circuit occurs along the positive direction of the $x$ axis. In the
Galilean limit ($c=\infty$), the velocity of the charge $\vec v_c$ can
be written as $\vec v_c=\vec V+\vec v_d$, where $\vec v_d$ is the
drift velocity of the charges} \footnote{{For a rigid and filiform
circuit at rest in the laboratory, the drift velocity $\vec v_d$  is
defined as the velocity of mobile charges when a steady or slowly
varying current flows. When the circuit moves inertially with
velocity $V$ along the positive direction of the common $x\equiv
x'$ axis, the drift velocity is defined in the moving reference frame
in which the circuit is at rest and is denoted by $\vec v'_d$. From
the above definition, it follows that $\vec v_d=\vec v'_d$, because
every phenomenon develops similarly in every inertial frame. In
other words: if we measure the drift velocity in a circuit in the
laboratory, we shall find a specific value $q$. If the {\em same}
circuit is in the moving inertial frame, and we measure the drift
velocity in this frame, we shall find the exact value $q$ measured
in the laboratory. Of course, this equivalence is true in special and
Galilean relativity. }}. { Then, Eq. (\ref{naturaldef}) assumes the
form:
\begin{equation}\label{galileanlimit}
\mathcal E =\oint_l \vec E\cdot\vec{dl}+ \oint_l  (\vec V\times\vec B)\cdot\vec{dl}+ \oint_l  (\vec v_d\times\vec B)\cdot\vec{dl},
\end{equation}
where all the line integrals are evaluated in the laboratory
reference frame. After some calculations \cite{ajp}, it can be
proved that the induced {\em emf}  is given by:
\begin{equation}\label{fr}
\mathcal E= -\frac{d}{dt}\int_S \vec B\cdot \hat n \,dS+\oint_l  (\vec v_d\times\vec B)\cdot\vec{dl}.
\end{equation}
The line integral is null for filiform circuits because the drift
velocity $\vec v_d$ is always parallel to $\vec{dl}$. Then, we get
the \fr. This rule has been obtained in the Galilean limit and for
inertially moving rigid and filiform circuits. Eq. (\ref{fr}) is also
valid in the reference frame of the circuit. Indeed, the \fr is Galileo
invariant, as it can be easily proved. In the Galilean limit $\vec
B'=\vec B$, $t'=t$, and $S'=S$, where the primed quantities refer to
the circuit's reference frame. Then $d\Phi'/dt'=d\Phi/dt$.}
\par
{The \fr is a piece of Galilean-Newtonian physics within the
Lorentz invariant theory of MLE. Approximations in the Galilean -
Newtonian limit can, of course, be used. However, an inescapable
condition is to discuss with the students the serious (physical and
epistemological) problems posed by the \fr. Moreover, the Galilean
limit of the law of electromagnetic induction is conceptually very
different from the Newtonian limit of relativistic dynamics. While
Newtonian dynamics can be interpreted causally, the \fr cannot
(see below). }
\par
{Therefore, instructors should underline that:
\begin{itemize}
\item The  \fr implies an improper use of the field concept,
because it describes what is going on in the closed circuit
with what happens -- at the same instant -- on an arbitrary
surface with the circuit as a contour. In this way, the  essential
feature of the field concept is lost: a field is a set of numbers
we specify in such a way that what happens at a point {\em of
the circuit} depends only on the numbers {\em at that point}.
We do not need to know anymore about what is happening at
other places ({\em on the surface with the circuit as a
contour}). The reader will recognize in this statement what
Feynman said in the quote at page \pageref{ft}, adapted to our
case.
\item The \fr cannot be interpreted causally because it relates
the physical quantity $\mathcal E$ defined on the line $l$ at
the instant $t$ to the values of the magnetic field $\vec B$
defined at all points of an arbitrary surface at the same time
$t$, thus implying the propagation of physical interaction with
infinite speed (see also the discussion of Eq. (\ref{roteint}) in
the next section).
\item It cannot say where the induced {\em emf} is localized
\cite{ajp}. To illustrate this point, instructors should discuss
the case (generally treated in textbooks) of a bar sliding on a
U-shaped conductive frame immersed in a constant and
uniform magnetic field. As shown in \cite{ajp}, the induced
{\em emf} is localized in the bar for both inertial reference
frames (the laboratory's and the bar's).
\item Frequently, it requires an {\em ad hoc} choice of the path
used as a contour of the integration surface \cite{scanlon}.
\item As shown by Blondel (1914) \cite{blondel}, it is falsified by
a clear-cut experiment \cite{ajp}.
\end{itemize}
In a study of electromagnetic induction understanding by first
years university students, Guisasola et al. found ``that most of the
students failed to distinguish between macroscopic levels
described in terms of fields and microscopic levels described in
terms of the actions of fields'' \cite{guisa}. According to the
authors, the \fr is a macroscopic description, while Lorentz's force
is microscopic. The definition of the induced {\em emf} given by
Eq. (\ref{naturaldef}) is a microscopic description. {If developed
coherently,  it leads to a microscopic theory
of electromagnetic induction.}
\par
{Teaching electromagnetic induction with a full microscopic
description will avoid the use and the pitfalls of the macroscopic
description of the \fr. Instructors have to make a choice depending
on their teaching context. If, as tradition, textbooks, and teaching
habits imply, the choice is the \fr, this choice should be
accompanied by a full discussion of its physical and
epistemological drawbacks. What should be avoided is speaking
of the \fr as the law of electromagnetic induction without any
critical discussion, which, by the way, would stimulate the
students' critical reasoning.}

\subsection{What to say about Maxwell's equations?}\label{what}
A choice is that of ignoring them. Giancoli does not even mention
Maxwell \cite{giancoli}. Another option is to write them in integral
form. Italian textbooks for high school widely adopt this choice
\cite[pp. 233-234]{paolo}, \cite[pp. 299-306]{parodi}, \cite[pp.
105-106]{bocci}. In the United States, we have encountered an
example in Halliday, Resnick, and Walker's book \cite[pp.
941-951]{halres}. Instead, Cutnell and Johnson do not mention
Maxwell's equations \cite{cj} \footnote{{We have not been able to
ascertain if these texts are considered in the United States or in
other English speaking countries as textbooks for high school {or for higher teaching levels}. We
know that their Italian translations are considered as textbooks for high school.}}.
\par
{Indeed, all textbooks  speak about two of
Maxwell's equations in integral form (or as a sum of finite terms of
the type $\vec E\cdot \vec {\Delta S}$ (Gauss law
\footnote{{Coulomb's law is written in terms of forces between
point charges; instead Gauss's law is written in terms of the
electric field, i.e., of a quantity defined at every point of the
considered surface. If this step is done without a clear
explanation, it would likely confuse the students because it
overlaps the action-at-a-distance and the field descriptions. }}) or
$\vec B\cdot \vec {\Delta S}$ (\fr), perhaps without labeling them as
Maxwell's equations.}
\par
Writing Maxwell's equations in integral form is conceptually
deceptive. For instance, consider the equation:
\begin{equation}\label{rote}
{\rm curl} \vec E=-\frac{\partial\vec B}{\partial t},
\end{equation}
and its integral form:
\begin{equation}\label{roteint}
\oint_l \vec E\cdot\vec {dl}= - \frac{d}{dt}\int_S \vec B\cdot\hat n\, dS.
\end{equation}
This equation relates what happens on the closed line $l$ at the
time $t$ to what happens, at the same time $t$, on an arbitrary
surface $S$ with the line $l$ as a contour. This equation cannot be
interpreted causally because physical interactions cannot
propagate at infinite speed. Eq. (\ref{roteint}) only establishes a
relation between quantities defined on the line with quantities on
the arbitrary surface chosen.
\par
Textbooks and teaching practices, in discussing Eq. (\ref{roteint}),
state that a varying magnetic field produces (causes) an electric
field; and, conversely, from the equation of the curl of the
magnetic field (or its integral form), they state that a varying
electric field produces (causes) a magnetic field. These statements
are untenable because the electric and magnetic fields are
produced (caused) by charges in motion. Therefore, the equations
of the curl of the electric and magnetic fields (or their integral
form) only establish a relation between these fields without any
causal connection between them. These issues are widely
discussed in \cite{ajp, hill}.
\par
{Instructors face a crossroads. Keeping on using Maxwell's
equations in integral form, explicitly (as Italian instructors,
following their textbooks, do) or follow a more challenging way
outlined in section \ref{maxeq} of the Appendix. In the first case,
instructors should explain the physical and epistemological
drawbacks of this choice}.

\section{Discussion and conclusions}\label{discon}

Instructors' resistance to proposed changes in teaching coming
from central or local institutions is well-studied in the literature.
Powell and Kusuma-Powell distinguish between ``technical and
``adaptive'' changes. Technical changes require informational
learning. Instead, adaptive changes ``call for transformational
learning or learning that requires us to rethink our deeply held
values, beliefs, assumptions, and even our professional identity.
Adaptive challenges are complex, and addressing them requires
patience and time \cite[p. 67]{pk}.'' Our proposal demands
instructors the disposition to:
\begin{enumerate}
\item[a] abandon the centuries-old tradition of presenting
electromagnetic phenomena following their chronological
development
\item[b] leave behind the epistemological stand according to
which physical laws must be induced only from experiment
\item[c] acknowledge the fundamental role played by the
abstraction and the hypothetical-deductive method
\item[d] recognize the necessity of some essential, microscopic
descriptions.

\end{enumerate}

{These features place our proposal in the ``adaptive changes'' category and could be evaluated as too radical to be implemented by instructors.}
We have presented an early version of our proposal to a group of
about twenty-five Italian instructors we meet periodically online.
{The major part of these instructors teach in the scientific Lyceum.
Positive reactions came from retired
instructors}. The negative reaction has been substantially based
on the following:
\begin{enumerate}
\item {The backwardness of the teaching context.}
\item The constraints of the programs of the Ministry of
Education and the local tendency to standardize the teaching
practices in all the classrooms, with the adoption of the same
textbooks, also in the view of preparing the students for the
final state exam.
\item The nonnecessity of teaching MLE Electromagnetism;
some Galilean approximation of standard courses is
sufficient.\label{galileo}
\item The mathematical difficulties.\label{math}
\end{enumerate}
The first two points are sadly founded. As explained
above, we cannot agree with point \ref{galileo} because we believe
that electromagnetic phenomena must be taught within the
conceptual framework of MLE electromagnetism. Instead, we have
thoroughly considered the last point (\ref{math}). Meanwhile, four
teachers in our group have agreed to the project of experimenting
with the present proposal in their classes. We are actively working with them on this project at a pace of about a meeting per month. }
\vskip2mm\par\noindent
{\bf Acknowledgments.} We warmly thank Maria Grazia Blumetti, Elena Failla, Andrea Farusi,   and Marco Litterio for their suggestions and commitment to experiment with this project.
\vskip4mm\par\noindent
\appendix\small
\counterwithin*{equation}{section}
\renewcommand\theequation{\thesection\arabic{equation}}
In this Appendix, instructors will find some development of topics
in the main text that could be used in favorable teaching
conditions or as instructors' background knowledge. {For instance, while section \ref{gravfield} belongs to the former group, the other two sections are primarily -- but not exclusively -- intended for the instructors'  background knowledge.}
\section{Gravitational field}\label{gravfield}
\begin{figure}[htb]
\centering{
\includegraphics[width=7cm]{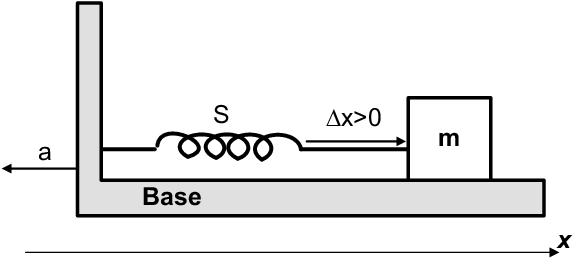}
}
\caption{\label{acce}  Working principle of an accelerometer.}
\end{figure}

 As we have seen in
section \ref{field}  of the main text, the gravitational field has the
dimensions of an acceleration. Accelerations can be measured
with an accelerometer in the accelerated reference frame (Fig.
\ref{acce}).
\par\noindent
The mass $m$ is connected to a rigid  base by the spring $S$;
ideally, it can slide on the base without friction. Suppose the base
is subjected to a constant acceleration to the left. In that case, the
spring is stretched, and its maximum extension $\Delta x$ is
related to the acceleration of the base by the equation:
\begin{equation}\label{accelerometr}
\overrightarrow{\Delta x}=-\frac{m}{k} \vec a,
\end{equation}
where $k$ is the spring's constant. The elongation of the spring
occurs along the opposite direction of the base's acceleration. If
the accelerometer is rotated 90 degrees to the right, it will find
itself in the vertical position. The spring elongates towards the
ground, owing to the effect of the gravitational field $\vec g$ on
the mass $m$: the accelerometer becomes a gravimeter. In this
situation, the accelerometer indicates an acceleration equal to
$-\vec g$ directed upwards \footnote{The gravitational field
measured on the Earth's surface depends on the latitude, also if
we suppose that the Earth's surface is spherical. In fact, in the
accelerated reference system centered at the Earth center and
rotating with the Earth, the component of the centripetal
acceleration perpendicular to the Earth's surface is equivalent to a
pseudo-gravitational field directed upwards. This
pseudo-gravitational field decreases the value of $g$ measured by
an accelerometer. In particular, the gravitational field is smaller at
the equator than the pole, as it can be easily proved.}.
\par If a laboratory --  with the accelerometer fixed in the vertical
position on a wall --  is free falling in a gravitational field, the
spring does not elongate because the (pseudo-gravitational) field
$-\vec g$ is canceled out by the acceleration $\vec g$ due to free
fall: the measured acceleration is null. {Since we have defined an inertial reference frame one in which the measured acceleration is null, it follows that a free-falling laboratory constitutes an inertial reference frame.} Moreover, this thought experiment
suggests that the effect of a gravitational field $\vec g$ on a mass
$m$ is equivalent to the effect of an acceleration field $-\vec g$ on
the same mass. Therefore, we can conclude that $m_g=m_i$,
where $m_g$ is the ``gravitational mass'' and $m_i$ is the ``inertial
mass'' which appears in the Newtonian equation $\vec F=m_i \vec
a$. In metric theories of gravitation, this property is assumed as
the ``weak equivalence principle'' \footnote{In special relativity, the
mass $m$ is no longer a measure of a body's inertia. Indeed, the
concept of inertial mass rests on using the equation $\vec F=m\vec a$, which is no longer valid in special relativity.}. {Since a
the free-falling laboratory is an inertial reference frame, a body that is left free will remain at rest or,
if endowed with an initial linear momentum, it will keep moving
uniformly along a straight line.} However, this is true only if the gravitational field is
uniform: in general, gravitational fields are not. Hence, the
previous statement is approximately verified, provided the
laboratory sizes are sufficiently small.
\begin{figure}[h]
\centering{
\includegraphics[width=3.0cm]{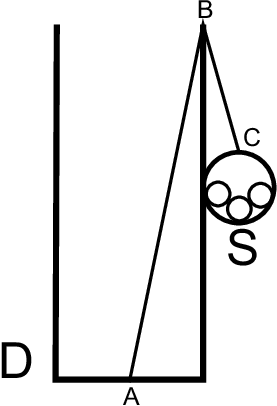}\\
}
\caption{D is a plastic cylinder; A-B-C is a narrow band of elastic rubber; S is a plastic ball or cylinder (made of two parts
that can be separated) containing a suitable number
of coins. This device can be quickly built using materials easily found at home.}\label{free-fall}
\end{figure}
\par\noindent
The qualitative features of a free-falling body can be demonstrated
in the classroom using the simple device shown in Fig.
\ref{free-fall} \footnote{This home made device has been
suggested to one of the author (G.G.) by Prof. Mauro Carfora.}.
The instructor should perform two experiments. Before doing each
experiment, the instructor illustrates what he will do and asks the
student what will happen. The first experiment lets the device fall
from the instructor's hand and is positioned from the ground at
the highest possible level. The ball containing the coins will be
retracted into the cylinder during free fall. The second experiment
consists in launching the cylinder toward the ceiling. The ball
containing the coins will be retracted into the cylinder already
during the ascent to the ceiling, thus demonstrating that the free
fall is the motion of a mass under the unique action of a
gravitational field. The discussion with the student will also
encompass the negligible effect of the atmosphere.
\par
Finally, it would be interesting to discuss with the students a
passage from a book by Galileo Galilei that reads \cite[pp.
63-64]{galilei}:
\begin{quote}\small
A large stone placed in a balance not only acquires additional
weight by having another stone placed upon it, but even by the
addition of a handful of hemp its weight is augmented six to ten
ounces according to the quantity of hemp. But if you tie the hemp
to the stone and allow them to fall freely from some height, do you
believe that the hemp will press down upon the stone and thus
accelerate its motion or do you think the motion will be retarded
by a partial upward pressure? One always feels the pressure upon
his shoulders when he prevents the motion of a load resting upon
him; but if one descends just as rapidly as the load would fall how
can it gravitate or press upon him? Do you not see that this would
be the same as trying to strike a man with a lance when he is
running away from you with a speed which is equal to, or even
greater, than that with which you are following him? You must
therefore conclude that, during free and natural fall, the small
stone does not press upon the larger and consequently does not
increase its weight as it does when at rest \footnote{Una gran
pietra messa nella bilancia non solamente acquista peso maggiore
col soprapporgli un'altra pietra, ma anco la giunta di un
pennecchio di stoppa la far\`a pesar pi\`u quelle sei o dieci once
che peser\`a la stoppa; ma se voi lascerete liberamente cader da
un'altezza la pietra legata con la stoppa, credete voi che nel moto
la stoppa graviti sopra la pietra, onde gli debba accelerar il suo
moto, o pur credete che ella la ritarder\`a, sostenendola in parte?
Sentiamo gravitarci su le spalle mentre vogliamo opporci al moto
che farebbe quel peso che ci sta addosso; ma se noi scendessimo
con quella velocit\`a che quel tal grave naturalmente scenderebbe,
in che modo volete che ci prema e graviti sopra? Non vedete che
questo sarebbe un voler ferir con la lancia colui che vi corre
innanzi con tanta velocit\`a, con quanta o con maggiore di quella
con la quale voi lo seguite? Concludete pertanto che nella libera e
naturale caduta la minor pietra non gravita sopra la maggiore, ed
in consequenza non le accresce peso, come fa nella quiete\cite[pp.
76-77]{galilei2}.}.
\end{quote}
\section{Maxwell's equations}\label{maxeq} In section  \ref{what} of the
main text, we have discussed using (at least two) Maxwell's
equations written in integral form: we have stressed this
treatment's physical and epistemological drawbacks.
{Generally speaking, the teaching contexts allow a variety of choices. In
the following, we will outline a more challenging path that could -- perhaps --
be followed in suitable conditions.}
\par
Instructors could write Maxwell's equations in a vacuum in their
differential form without, however, specifying the expression of
the divergence and curl operators:
\begin{eqnarray}
{\rm div}\: \vec E & = & {{\rho} \over {\varepsilon_0 }} \label{dive}\\
{\rm curl}\: \vec E & = & -{{\partial \vec B} \over {\partial t}}
\label{rotee}\\ {\rm div}\: \vec B & = & 0 \label{divb}\\ {\rm curl}\: \vec
B & = & \mu_0  \left(\vec J + \varepsilon_0 {{\partial \vec E} \over
{\partial t}}    \right), \label{rotb} \end{eqnarray} and comment on
them in the following way: \begin{enumerate} \item The operator
divergence and curl operate on the spatial variations of the vector
to which they are applied. \item These equations relate the {\em
sources} $\rho$  (charge density) and $\vec J$ (current density) to
the electric field $\vec E$ and to the magnetic field $\vec B$. \item
The first equation (\ref{dive}) states that the operator divergence
applied to the electric field  $\vec E$ yields $\rho/\varepsilon_0$.
\item The third equation (\ref{divb})  says that the divergence of
the magnetic field is always null. This result implies that the
magnetic field has no sources similar to the charge density for
the electric field. Indeed, the magnetic field sources are the
currents' densities, i.e., charges in motion. \item The second
equation (\ref{rotee}) connects spatial variations of the electric
field $\vec E$ to the time variation of the magnetic field $\vec B$.
\item The fourth equation (\ref{rotb}) connects spatial variations of
the magnetic field to its source $\vec J$ and to the time
variation of the electric field. \item Given the sources $\rho$ and
$\vec J$, the physical dimensions of the electric and magnetic
fields remain undeterminate, together with those of the two
constants $\varepsilon_0$ and $\mu_0$. \item The assumption
of the Lorentz force $\vec F=q(\vec E+\vec v_q\times\vec B)$
gives physical dimensions to the two fields and the two
constants. \item The value of the two constants $\varepsilon_0$
and $\mu_0$ must be determined experimentally. \item
Maxwell's equations (\ref{dive} -- \ref{rotb}) describe all
electromagnetic phenomena observed in a vacuum. Further
assumptions must be made for describing electromagnetic
phenomena in a material. \item The solutions of Maxwell'e
equations describe how electromagnetic signals produced by
the sources propagate. In a vacuum, their propagation velocity
is $c=1/\sqrt{\varepsilon_0\mu_0}$. \item If the sources do not
depend on time, Maxwell's equations describe electrostatic
phenomena. \item In 1888, Hertz demonstrated that
electromagnetic waves reflect, refract, and diffract as light
waves; they are also polarized. Light and electromagnetic waves
obey the same equations. Hence, light can be described as an
electromagnetic wave. \item Special relativity shows that $c$ is
a limit speed. \item As for the two constants, their numerical
values are obtained by putting $\mu_0=4\pi\times 10^{-7} {\rm
NA^{-2}}$ and deducing $\varepsilon_0$ from the formula
yielding the light velocity in a vacuum determined
experimentally. \item Instructors should add that Maxwell's
equations written for a magnetic material assume that their
magnetic properties are due to currents circulating in the
material (Amp\`re's currents). Indeed, a sound explanation of
magnetic properties requires a quantum mechanical treatment.
\end{enumerate}
\section{Electromagnetic induction}\label{emisec} In section \ref{emi}
of the main text, we have seen how the \fr is only a calculation
shortcut and pointed out that the law of electromagnetic induction
is founded on the definition of the induced {\em emf} as:
\begin{equation}\label{naturaldefSM}
\mathcal E=\oint_l (\vec E+\vec v_c\times\vec B)\cdot\vec{dl}= \oint_l \vec E\cdot\vec{dl}+ \oint_l  (\vec v_c\times\vec B)\cdot\vec{dl}.
\end{equation}
We have observed that this equation is local and that also its
solution must be local. Consequently, both equations can be
interpreted causally. In the following, we develop some
calculations that should be part of the background knowledge of
instructors on this topic.
\par
Within the description of MLE in terms of the electromagnetic
potentials, the general expression of the electric field is given by:
\begin{equation}\label{pot} \vec E=
-\nabla\varphi-\frac{\partial \vec A}{\partial t}, \end{equation}
where $\varphi$ and $\vec A$ are the scalar and the vector
potential. Consequently, Eq.  (\ref{naturaldefSM}) assumes the
form:
\begin{equation}\label{natural2}
\mathcal E =
\oint\left[\left(-\nabla\varphi- \frac{\partial \vec A}{\partial t}\right) + (\vec
v_c \times \vec B)\right]\cdot \vec{dl}= \oint_l \left[\left(-
\frac{\partial \vec A}{\partial t}\right) + (\vec v_c \times \vec
B)\right]\cdot \vec{dl}, \end{equation}
because the line integral $\oint_l {\rm grad}\,\varphi\cdot\vec
{dl}=0$.
\par
If we want to get the \fr, we must start again from the equation
(\ref{naturaldefSM}), and write, in the reference frame of the
laboratory:
\begin{eqnarray}\label{onlyone}
\mathcal E=  \oint _{l}^{}{\vec E\,\cdot\, \vec{dl}} + \oint_l (\vec v_c\times\vec B) \cdot \vec{dl}&= &\int_{S} {\rm curl}\, \vec E \,
\cdot \, \hat n \, dS + \oint_l (\vec v_c\times\vec B) \cdot \vec{dl}\nonumber\\
&&\\
&= &- \int_{S} {{\partial \vec B} \over
{\partial t}} \, \cdot \, \hat n \, dS + \oint_l (\vec v_c\times\vec B) \cdot \vec{dl},\nonumber
\end{eqnarray}
where $S$ is any {\em arbitrary} surface that has the integration
line $l$ as a contour. For every vector field with mull divergence
\cite[pp. 10 - 11]{andy}):
\begin{equation}\label{identity}
\int_S \frac{\partial \vec B}{\partial t}\cdot  \hat n \,dS=
\frac{d}{dt}\int_S \vec B \cdot \hat n \,dS+\oint_l (\vec v_l\times \vec B)\cdot\vec{dl},
\end{equation}
where $\vec v_l$,  the velocity of the line element $dl$,  can be
different for each line element. Therefore,  equation (\ref{onlyone})
becomes:
\begin{equation}\label{quasiflusso}
{\mathcal E}=- \frac{d\Phi}{dt}- \oint_l(\vec v_l\times \vec B)\cdot\vec{dl} +\oint_l (\vec v_c \times \vec B)\cdot \vec{dl}.
\end{equation}
In the case of a {\em rigid, filiform  circuit} moving with velocity
$V$ along the positive direction of the common $x'\equiv x$ axis,
this equation becomes:
\begin{equation}\label{quasiflussorigido}
{\mathcal E}=- \frac{d\Phi}{dt}- \oint_l(\vec V\times \vec B)\cdot\vec{dl} +\oint_l (\vec v_c \times \vec B)\cdot \vec{dl}.
\end{equation}
We can write $\vec v_c= \vec V +\vec v_d$ in the Galilean limit
($c=\infty$). Then, finally:
\begin{equation}\label{quasiflux2}
\mathcal E=-\frac{d\Phi}{dt}
+\oint_l (\vec v_{d}\times \vec B)\cdot\vec dl= -\frac{d\Phi}{dt},
\end{equation}
i.e., the ``flux rule'' (the line integral is null because for every line
element, $\vec v_d$ is parallel to $\vec {dl}$).
\vskip4mm\par

\end{document}